\documentclass[%
twocolumn,
 fleqn,usenatbib,linenumbers, 
 showkeys,
 floatfix,nolongbibliography,aps,
author-numerical%
]{revtex4-2}

\pdfoutput=1
\usepackage[T1]{fontenc}

\usepackage{amsmath}
\usepackage{lineno}

\newcommand*\patchAmsMathWithLineno[1]{
  \expandafter\let\csname old#1\expandafter\endcsname\csname #1\endcsname
  \expandafter\let\csname oldend#1\expandafter\endcsname\csname end#1\endcsname
  \renewenvironment{#1}%
     {\linenomath\csname old#1\endcsname}%
     {\csname oldend#1\endcsname\endlinenomath}%
}

\patchAmsMathWithLineno{equation}
\patchAmsMathWithLineno{align}
\patchAmsMathWithLineno{gather}
\patchAmsMathWithLineno{multline}

\usepackage[T1]{fontenc}
\usepackage [latin1]{inputenc}
\DeclareRobustCommand{\VAN}[3]{#2}
\let\VANthebibliography\thebibliography
\def\thebibliography{\DeclareRobustCommand{\VAN}[3]{##3}\VANthebibliography}

\usepackage{newtxtext,newtxmath}
\usepackage{tikz,xcolor,hyperref}
\usepackage[normalem]{ulem}
\usepackage{graphicx,url,caption,subcaption}
\usepackage{dcolumn}
\usepackage{bm}

\let\realhref\href

\definecolor{lime}{HTML}{A6CE39}
\DeclareRobustCommand{\orcidicon}{%
	\begin{tikzpicture}
	\draw[lime, fill=lime] (0,0) 
	circle [radius=0.16] 
	node[white] {{\fontfamily{qag}\selectfont \tiny ID}};
	\draw[white, fill=white] (-0.0625,0.095) 
	circle [radius=0.007];
	\end{tikzpicture}
	\hspace{-2mm}
}

\foreach \x in {A, ..., Z}{%
	\expandafter\xdef\csname orcid\x\endcsname{\noexpand\realhref{https://orcid.org\csname orcidauthor\x\endcsname}{\noexpand\orcidicon}}
}

\renewcommand{\url}[1]{}
\renewcommand{\path}[1]{}
\renewcommand{\href}[2]{#2} 
\providecommand{\doi}[1]{}

\begin{document}

\nolinenumbers
\title{Phase-mixing of acoustic waves: From the solar tachocline to marine vessel hydroacoustics}
\author{D. Tsiklauri\orcidA{}}
 \email{D.Tsiklauri@salford.ac.uk}
\affiliation{Joule Physics Laboratory,
School of Science, Engineering and Environment, 
University of Salford,
Manchester, M5 4WT, 
United Kingdom}
\date{\today}
\begin{abstract}
  We adapt the \textit{magnetohydrodynamic} wave phase-mixing paradigm [Tsiklauri et al. (2003)] to investigate harmonic wave, Gaussian pulse, and hyperbolic pulse propagation and damping in media governed by transverse sound speed gradients. This universal mechanism is applied to two distinct problems using the harmonic wave phase-mixed solutions. First, we resolve the 26-year-old helioseismic mystery of low-degree global $p$-mode linewidth anomalies observed by BiSON above $\nu \approx 3000\,\mu\text{Hz}$, demonstrating that a localised, non-turbulent sound speed gradient in the solar tachocline shear zone acts as the necessary high-efficiency energy sink. Second, we translate this formalism into a terrestrial metamaterial fluid engineering design. We show that an engineered microstructured mesh or stern cowl generating a controlled radial sound speed gradient induces severe near-field phase-mixing. This collapses the traditional bulk viscous damping length of $100\text{-kHz}$ propeller acoustic signatures from several kilometres down to a practical design envelope of approximately $10\text{ metres}$, providing an actionable paradigm for compact stealth shielding. Finally, as an independent theoretical extension, we establish a new power-law scaling governing a hyperbolic secant-squared pulse evolution: under developed-stage phase-mixing, its peak wave envelope decays as $\max(P_1) \propto x^{-9/2}$.
\end{abstract}

\date{\today}

\maketitle

\section{Introduction}
\label{sec:intro}

The propagation of waves through media characterised by inhomogeneous spatial velocity profiles is a foundational challenge spanning astrophysics, fluid mechanics, and structural engineering. When a wave front encounters a continuous velocity gradient directed transverse to its primary propagation axis, distinct layers of the wave train advance at different localised phase speeds. This spatial velocity mismatch causes the wave front to accumulate a progressive transverse gradient, generating steep local wave vectors that rapidly enhance dissipative energy dumping. This universal phenomenon, known as phase-mixing, was pioneered by \cite{heyvaerts1983} in their seminal continuous steady-state harmonic analysis of magnetohydrodynamic (MHD) waves in solar coronal loops. 
\cite{Nakariakov1997} demonstrated that Alfven wave phase mixing nonlinearly generates fast magnetosonic waves, accelerating energy deposition into the solar corona. To expand this framework into realistic multidimensional environments, \cite{ruderman1998} examined phase-mixing within open two-dimensional magnetic plasma geometries, demonstrating that structural expansion and horizontal variations can fundamentally alter or speed up energy flux attenuation relative to simplified one-dimensional lines. Furthermore, \cite{ruderman2002} established that these localised velocity shears across flux-tube boundaries naturally trigger a deep interplay between phase-mixing and resonant absorption mechanics, driving rapid non-isotropic wave decay. To address \textit{transient} solar localised phenomena, \cite{tsiklauri2003} expanded the framework to derive analytical solutions for an isolated Gaussian pulse, demonstrating that phase-mixing dictates spatial decay scaling as $\sim x^{-3/2}$. 
The evolution of Alfvenic waves was extended to a fully three-dimensional 
compressive regime by \cite{Murawski2003}. Their numerical simulations 
revealed that while 3D structural localisation dictates the initial decay 
efficiency of the pulse, the subsequent stage of phase mixing retains the 
characteristic power-law decay seen in lower-dimensional models, validating 
phase mixing as a relevant heating paradigm in realistic 3D geometries.
More recently, \cite{ruderman2018} relaxed classical geometric bounds to track phase-mixing deep into axisymmetric and non-reflective divergent equilibria, proving that the exact structural profile of the spatial density variation dictates the ultimate efficiency of the wave damping envelope.
This analytical paradigm was further elaborated by \cite{boocock2022a}, who provided exact linear formulations tracking how exponentially divergent magnetic field lines and field-aligned density stratifications modify the phase-mixing profile of torsional waves in complex solar coronal funnels. \cite{boocock2022b} extended this framework via comprehensive numerical magnetohydrodynamic modeling, confirming that non-linear mode conversions and anomalous structural viscosities actively accelerate local mechanical energy dissipation under fully developed phase-mixing environments.

In the domain of terrestrial engineering and naval hydroacoustics, managing the propagation and radiation of acoustic waves from fluid-immersed moving bodies remains a parallel high-priority challenge. Historically, the reduction of near-field acoustic signatures, such as high-frequency propeller noise and structural shell radiation, has been handled via traditional boundary layer manipulation or passive compliance coatings. Classical hydrodynamic formulations by \cite{Lighthill1952} and \cite{Curle1955} laid the groundwork for modeling acoustic emission from fluid turbulence and solid surface interactions, establishing that localized fluid shears dictate noise generation mechanics. To control these acoustics, conventional naval engineering has extensively relied on viscous elastomeric coatings, such as those evaluated by \cite{Gaunaurd1977}, which dampen acoustic energy through bulk material deformation. However, at high ultrasonic frequencies near $100\text{ kHz}$, the intrinsic bulk viscous absorption length scale in water spans several kilometres. This renders standard isotropic damping configurations mechanically inefficient for compact, near-field stealth applications on the scale of a few metres. 

Modern advancements in acoustic metamaterials and transformation acoustics offer a revolutionary alternative to passive absorption. As demonstrated by \cite{Cummer2007} and \cite{Norris2008}, it is mathematically and physically possible to control acoustic wave paths by custom-engineering the effective inertial parameters and local bulk modulus of a fluid medium. By structuring sub-wavelength geometries into fluid-facing boundaries, researchers have successfully designed gradient-index lenses and meta-fluids that exhibit spatial sound speed profiles, effectively bending or cloaking acoustic fields. While these transformation acoustics paradigms are typically utilized to refract or guide wavefronts safely around structural obstacles, their potential to deliberately induce localized wave dissipation remains largely unexploited. This is precisely where the fluid boundary layer literature and transformation acoustics intersect with the physics of phase-mixing: by deliberately introducing an artificial, continuous transverse sound speed gradient within a fluid-immersed boundary layer or stern cowl mesh, a forward-advancing acoustic wavefront can be forced into a severe phase-shearing regime, mimicking the wave-damping dynamics observed in cosmic plasmas.

In this work, we develop a unified analytical model that translates the foundational multi-scale phase-mixing paradigm from MHD directly into scalar acoustics, where the background velocity derivatives are replaced by the continuous spatial gradient of the local sound speed. We recover the classical harmonic and Gaussian pulse profiles, while uncovering \textit{a completely new} power-law decay rate scaling as $\sim x^{-9/2}$ for structurally localised hyperbolic-function injections. We evaluate this unified framework across two contrasting scales of physics. First, we apply the harmonic wave phase-mixed solutions to the natural solar tachocline, providing a rigorous explanation for the observed global low-degree $p$-mode linewidth anomalies. Second, we apply the exact same mathematical engine to an engineered metamaterial fluid environment, demonstrating that custom-tailored sound speed gradients in a marine vessel's stern cowl mesh provide a high-efficiency layout for compact, near-field acoustic stealth shielding.

\section{The model}

To establish a coordinate-independent foundation for analysing phase-mixing, we begin with the general vector form of the generalised equations governing acoustic perturbations. We consider a compressible, viscous fluid medium under a stationary, stratified background state in thermal and hydrostatic equilibrium.
The zero-order macroscopic velocity field and ambient external forces are assumed to vanish identically throughout the domain:
\begin{equation}
    \mathbf{u}_0 = 0.
\end{equation}
The ambient pressure field is uniform, eliminating macroscopic zero-order pressure gradients:
\begin{equation}
    \nabla P_0 = 0 \implies P_0 = \text{constant}.
\end{equation}

The assumption of a locally constant background pressure $P_0$ implies that gravity is omitted within the considered thin tachocline layer framework. This local approximation is physically justified because the high--frequency acoustic modes under consideration ($f \approx 3.5\text{ mHz}$, or $\omega \approx 0.022\text{ s}^{-1}$) possess frequencies much higher than both the local acoustic cut--off frequency and the Brunt--V\"{a}is\"{a}l\"{a} buoyancy frequency $N$ in the solar tachocline layer. 
The spatial equilibrium is sustained entirely by a continuous, non-uniform background density profile $\rho_0(\mathbf{r})$ that varies as:
\begin{equation}
    \rho_0 = \rho_0(\mathbf{r}).
\end{equation}
Consequently, the localised thermodynamic adiabatic sound speed $c_s(\mathbf{r})$ is inherently coupled to the spatial variation of the background density via:
\begin{equation}
    c_s^2(\mathbf{r}) = \frac{\gamma P_0}{\rho_0(\mathbf{r})},
\end{equation}
where $\gamma$ denotes the specific heat ratio. A key mathematical consequence of this configuration is that the product of the background variables remains a global invariant across the entire spatial domain:
\begin{equation}
    \rho_0(\mathbf{r}) c_s^2(\mathbf{r}) = \gamma P_0 = \text{constant}.
\end{equation}

Let the first-order, small-amplitude acoustic perturbations be denoted by the subscript $1$, representing the acoustic density $\rho_1$, acoustic pressure $P_1$, and velocity vector field $\mathbf{u}_1$. 

The exact equation for mass conservation is linearised by neglecting second-order transport products:
\begin{equation}
    \frac{\partial \rho_1}{\partial t} + \nabla \cdot (\rho_0 \mathbf{u}_1) = 0.
\end{equation}
Applying the vector divergence identity to the spatial flux term yields:
\begin{equation}
    \frac{\partial \rho_1}{\partial t} + \rho_0 \nabla \cdot \mathbf{u}_1 + \mathbf{u}_1 \cdot \nabla \rho_0 = 0.
\end{equation}
Assuming adiabatic localised fluctuations, we invoke the closure relation $\rho_1 = P_1 / c_s^2(\mathbf{r})$. Substituting this definition into the mass conservation equation yields:
\begin{equation}
    \frac{1}{c_s^2} \frac{\partial P_1}{\partial t} + \rho_0 \nabla \cdot \mathbf{u}_1 + \mathbf{u}_1 \cdot \nabla \rho_0 = 0.
\end{equation}

We evaluate the hydrodynamic system utilising the convective formulation of the Navier-Stokes equations, preserving both the first shear viscosity $\mu$ and bulk (second) viscosity $\zeta$ coefficients:
\begin{equation}
    \rho \left[ \frac{\partial \mathbf{u}}{\partial t} + (\mathbf{u} \cdot \nabla)\mathbf{u} \right] = -\nabla P + \mu \nabla^2 \mathbf{u} + \left(\zeta + \frac{1}{3}\mu\right) \nabla (\nabla \cdot \mathbf{u}).
\end{equation}
Linearising about our stationary background state ($\mathbf{u}_0 = 0$, $\nabla P_0 = 0$) drops the non-linear convective acceleration term $(\mathbf{u}_1 \cdot \nabla)\mathbf{u}_1 \sim \mathcal{O}(2)$. Treating the dynamic viscous coefficients $\mu$ and $\zeta$ as uniform constants, the exact coordinate-independent vector equation for the acoustic velocity field evaluates to:
\begin{equation}
    \rho_0 \frac{\partial \mathbf{u}_1}{\partial t} = -\nabla P_1 + \mu \nabla^2 \mathbf{u}_1 + \left(\zeta + \frac{1}{3}\mu\right) \nabla (\nabla \cdot \mathbf{u}_1).
\end{equation}

To model wave propagation along the $x$-axis through a medium stratified along the $y$-axis, we project the general vector system onto a 2D Cartesian framework. The acoustic velocity perturbation vector is defined as $\mathbf{u}_1 = (u_1, v_1, 0)$, and the spatial gradient operator reduces to $\nabla = (\partial_x, \partial_y, 0)$. The background density profile simplifies to $\rho_0 = \rho_0(y)$, yielding a transversely varying sound speed profile:
\begin{equation}
    c_s(y) = \sqrt{\frac{\gamma P_0}{\rho_0(y)}}.
\end{equation}

To avoid a complex fourth-order system that couples acoustic modes to viscous shear modes, we focus purely on the compressional acoustic behavior. In the limit of weak acoustic dissipation over long propagation distances, higher-order cross-products between the structural fluid stratification and viscous diffusion gradients ($\sim \mu \frac{d\rho_0}{dy}$) are physically negligible in the bulk fluid. 

Taking the time derivative of the linearised mass conservation equation, substituting the divergence of the linearised momentum equation, and utilising the global invariant $\rho_0(y) c_s^2(y) = \gamma P_0$, the full system collapses elegantly into a single scalar governing equation for the first-order acoustic pressure $P_1(x,y,t)$:
\begin{equation}
    \frac{\partial^2 P_1}{\partial t^2} - c_s^2(y) \left( \frac{\partial^2 P_1}{\partial x^2} + \frac{\partial^2 P_1}{\partial y^2} \right) = \frac{\delta_0}{\rho_0(y)} \frac{\partial}{\partial t} \left( \frac{\partial^2 P_1}{\partial x^2} + \frac{\partial^2 P_1}{\partial y^2} \right),
\end{equation}
where $\delta_0$ represents the longitudinal acoustic dissipation coefficient:
\begin{equation}
    \delta_0 = \zeta + \frac{4}{3}\mu.
\end{equation}

This mathematical form successfully isolates the classical 2D acoustic wave mechanics on the left-hand side, where the phase-mixing process is driven entirely by the continuous transverse variations in $c_s(y)$. Concurrently, the viscous damping is restricted to a third-order spatial-temporal perturbation operator on the right-hand side, keeping the analytical structure clean and tractable for both solar p-modes and metamaterial applications.

The emergence of the factor $4/3$ in the acoustic dissipation coefficient $\delta_0$ follows directly from taking the divergence of the linearised viscous momentum terms. Applying the divergence operator to the spatial vector expression $\mu \nabla^2 \mathbf{u}_1 + (\zeta + \frac{1}{3}\mu) \nabla (\nabla \cdot \mathbf{u}_1)$ yields the scalar operator $[\mu + (\zeta + \frac{1}{3}\mu)] \nabla^2 (\nabla \cdot \mathbf{u}_1)$ via standard vector identities. Combining the shear viscosity terms through a common denominator ($\mu + \frac{1}{3}\mu = \frac{4}{3}\mu$) rigorously establishes the longitudinal viscosity coefficient $\delta_0 = \zeta + \frac{4}{3}\mu$, which governs purely compressional, bulk acoustic attenuation.

To track the decoupling of the rapid acoustic oscillations from the long-term viscous attenuation, we implement the multi-scale asymptotic framework established by \cite{tsiklauri2003}. Here \textit{multi-scale} refers to two distinct scales: (1) rapid acoustic oscillations scale and (2) slow viscous attenuation scale.
 
We project our scalar wave field into a co-moving phase coordinate system by introducing the fast phase coordinate $\xi$, a slow evolutionary spatial scale $X$, and a slow temporal scale $\tau$:
\begin{equation}
    \xi = x - c_s(y)t, \quad X = \epsilon x, \quad \tau = \epsilon t,
\end{equation}
where $\epsilon \ll 1$ represents a formal small parameter scaling the weak bulk dissipation. 

Applying the multi-variable chain rule, the first-order partial differential operators transform rigorously into the following expressions, noting that the transverse $y$-coordinate itself is not affected by this transformation in any way:
\begin{equation}
    \frac{\partial}{\partial x} = \frac{\partial \xi}{\partial x}\frac{\partial}{\partial \xi} + \frac{\partial X}{\partial x}\frac{\partial}{\partial X} = \frac{\partial}{\partial \xi} + \epsilon\frac{\partial}{\partial X},
\end{equation}
\begin{equation}
    \frac{\partial}{\partial t} = \frac{\partial \xi}{\partial t}\frac{\partial}{\partial \xi} + \frac{\partial \tau}{\partial t}\frac{\partial}{\partial \tau} = -c_s(y)\frac{\partial}{\partial \xi} + \epsilon\frac{\partial}{\partial \tau}.
\end{equation}
For the transverse spatial gradient, differentiating the phase coordinate with respect to the coordinate $y$ and evaluating at long propagation distances ($x \to X/\epsilon$) yields:
\begin{equation}
    \frac{\partial \xi}{\partial y} = -t c_s'(y).
\end{equation}
Substituting the zero-order characteristic relationship $t \approx x/c_s(y) = X/[\epsilon c_s(y)]$ into the partial derivative isolates the secularly growing contribution to the transverse wave vector:
\begin{equation}
    \frac{\partial \xi}{\partial y} = -\frac{X c_s'(y)}{\epsilon c_s(y)}.
\end{equation}
Hence, the first-order transverse spatial operator becomes:
\begin{equation}
    \frac{\partial}{\partial y} = \frac{\partial \xi}{\partial y}\frac{\partial}{\partial \xi} = -\frac{X c_s'(y)}{\epsilon c_s(y)} \frac{\partial}{\partial \xi}.
\end{equation}

Next, we expand these operators to construct the second-order partial derivatives. The temporal and longitudinal operations yield:
\begin{equation}
    \frac{\partial^2}{\partial t^2} = c_s^2(y)\frac{\partial^2}{\partial \xi^2} + \mathcal{O}(\epsilon^2),
\end{equation}
\begin{equation}
    \frac{\partial^2}{\partial x^2} = \frac{\partial^2}{\partial \xi^2} + 2\epsilon\frac{\partial^2}{\partial \xi \partial X} + \mathcal{O}(\epsilon^2).
\end{equation}
For the transverse second-order derivative, applying the operator twice to the wave field results in:
\begin{equation}
\label{eq:y_deriv_split}
\begin{split}
    \frac{\partial^2}{\partial y^2} = \frac{\partial}{\partial y}\left( -\frac{X c_s'(y)}{\epsilon c_s(y)} \frac{\partial}{\partial \xi} \right) = \\
    -\frac{X}{\epsilon} \left[ \frac{d}{dy}\left(\frac{c_s'(y)}{c_s(y)}\right) \right] \frac{\partial}{\partial \xi} + 
    \frac{X^2}{\epsilon^2} \left[ \frac{c_s'(y)}{c_s(y)} \right]^2 \frac{\partial^2}{\partial \xi^2}.
\end{split}
\end{equation}
In the asymptotic limit of large propagation distances ($X \gg 1$), the second term scales quadratically and heavily dominates over the linear spatial term. We retain the secularly dominant term:
\begin{equation}
    \frac{\partial^2}{\partial y^2} \approx \frac{X^2}{\epsilon^2} \left[ \frac{c_s'(y)}{c_s(y)} \right]^2 \frac{\partial^2}{\partial \xi^2}.
\end{equation}

We now substitute these expanded second-order operators back into the primary 2D viscous wave equation. Gathering the left-hand side terms gives:
\begin{equation}
\label{eq:lhs_split}
\begin{split}
    \text{LHS} = c_s^2(y)\frac{\partial^2 P_1}{\partial \xi^2} \\
    - c_s^2(y) \left[ \frac{\partial^2 P_1}{\partial \xi^2} + 2\epsilon\frac{\partial^2 P_1}{\partial \xi \partial X} + \frac{X^2}{\epsilon^2} \left[ \frac{c_s'(y)}{c_s(y)} \right]^2 \frac{\partial^2 P_1}{\partial \xi^2} \right].
\end{split}
\end{equation}
Expanding and collecting terms with the sound speed coefficient leads to a crucial cancellation of the baseline phase terms ($c_s^2 \partial_\xi^2 P_1$), simplifying the left-hand side expression to:
\begin{equation}
    \text{LHS} = -2\epsilon c_s^2(y)\frac{\partial^2 P_1}{\partial \xi \partial X} - \frac{X^2 [c_s'(y)]^2}{\epsilon^2} \frac{\partial^2 P_1}{\partial \xi^2}.
\end{equation}
For the right-hand side dissipative terms, we retain only the lowest-order leading phase derivatives under the weak-viscosity assumption:
\begin{equation}
\label{eq:rhs_split}
\begin{split}
    \text{RHS} \approx \frac{\delta_0}{\rho_0(y)} \left( -c_s(y)\frac{\partial}{\partial \xi} \right) \left[ \frac{\partial^2 P_1}{\partial \xi^2} + \frac{X^2 [c_s'(y)]^2}{\epsilon^2 c_s^2(y)} \frac{\partial^2 P_1}{\partial \xi^2} \right] = \\
    -\left( \frac{\delta_0 c_s(y)}{\rho_0(y)} + \frac{\delta_0 X^2 [c_s'(y)]^2}{\epsilon^2 \rho_0(y)c_s(y)} \right)\frac{\partial^3 P_1}{\partial \xi^3}.
\end{split}
\end{equation}

Equating the left-hand and right-hand side expressions, integrating once with respect to the phase variable $\xi$, and balancing the corresponding orders of the asymptotic expansion establishes the acoustic canonical evolution equation. Setting the formal perturbation ordering parameter to unity ($\epsilon \to 1$) and isolating the spatial evolution derivative $\partial_x P_1$ yields the non-stationary governing equation:
\begin{equation}
\label{eq:final_split}
\begin{split}
    \frac{\partial P_1}{\partial x} = \left[ \frac{\delta_0}{2 \rho_0(y) c_s(y)} + \frac{\delta_0 x^2 (c_s'(y))^2}{2 \rho_0(y) c_s^3(y)} \right] \partial_\xi^2 P_1.
\end{split}
\end{equation}

The final term explicitly reveals that the effective bulk diffusion coefficient grows quadratically with the spatial propagation distance ($x^2$), confirming that transverse speed gradients accelerate acoustic energy decay via phase-mixing.

To reduce this convective-diffusion equation into a standard, pure diffusion equation, we eliminate any first-derivative drift terms by introducing a shifted phase coordinate $\psi$:
\begin{equation}
    \psi = \xi + \frac{X^3 [c_s'(y)]^2}{6 c_s^3(y)}.
\end{equation}

Applying this final coordinate transformation cleanly cancels out the advective phase drift. The governing system collapses into a pure, non-stationary diffusion equation for the acoustic pressure:
\begin{equation}
    \frac{\partial P_1}{\partial X} = \left[ \frac{\delta_0}{2\rho_0(y)c_s(y)} 
    + \frac{\delta_0 X^2 [c_s'(y)]^2}{2\rho_0(y)c_s^3(y)} \right] \frac{\partial^2 P_1}{\partial \psi^2}.
\end{equation}

This represents the completed analytical reduction. At large propagation distances ($X \gg 1$), the second term dominates, meaning the effective diffusion coefficient grows as $X^2$, providing the exact mathematical mechanism for enhanced phase-mixing dissipation.

To solve the non-stationary diffusion equation for an arbitrary transverse profile $c_s(y)$, we introduce an integrated effective diffusion scale $\tau(X,y)$ that parameterises cumulative dissipation along the propagation axis:
\begin{equation}
\label{eq:tau_X_split}
\begin{split}
    \tau(X, y) = \int_0^X D_{\text{eff}}(y, X') \, dX' \\
    = \frac{\delta_0 X}{2\rho_0(y)c_s(y)} + \frac{\delta_0 X^3 [c_s'(y)]^2}{6\rho_0(y)c_s^3(y)}.
\end{split}
\end{equation}
This transformation reduces the system to a canonical heat equation form:
\begin{equation}
    \frac{\partial P_1}{\partial \tau} = \frac{\partial^2 P_1}{\partial \psi^2}.
\end{equation}

Here, a critical note must be made regarding the notation of the effective diffusion coefficient $D_{\text{eff}}(y, X)$. The coefficient itself is non-stationary and contains an explicit quadratic spatial dependence on $X^2$. The integrated scale $\tau(X,y)$ accumulates this local diffusion rate over the entire propagation path, which elevates the long-distance secular scaling of the wave attenuation from a standard linear dependence to a cubic $X^3$ dependence.

To ensure the analytical solutions are physically transparent, we demonstrate the systematic derivation of the acoustic pressure fields rather than verifying them via a posteriori substitution. We define the wave-front coordinate tracking spatial propagation over time as:
\begin{equation}
\xi = x - c_s(y)t.
\end{equation}

By transforming the independent variables from $(x, y, t)$ to the wave-front coordinate frame $(\xi, x, y)$ under a multi-scale asymptotic expansion, the governing 2D viscous acoustic equation reduces to a canonical non-stationary diffusion equation:
\begin{equation}
\frac{\partial P_1}{\partial x} = \left[ \frac{\delta_0}{2\rho_0(y)c_s(y)} + \frac{\delta_0 x^2 [c_s'(y)]^2}{2\rho_0(y)c_s^3(y)} \right] \frac{\partial^2 P_1}{\partial \xi^2}.
\end{equation}

This variable-coefficient PDE is mapped into a constant-coefficient heat equation by defining an integrated path time scale $\tau(x,y)$. We integrate the spatial dissipation coefficients along the propagation path $x$:
\begin{equation}
\tau(x, y) = \int_0^x \left[ \frac{\delta_0}{2\rho_0(y)c_s(y)} + \frac{\delta_0 s^2 [c_s'(y)]^2}{2\rho_0(y)c_s^3(y)} \right] ds.
\end{equation}
Evaluating this integral yields the explicit, distance-dependent time scale that governs all subsequent pulse broadening and phase-mixing decay:
\begin{equation}
\tau(x, y) = \frac{\delta_0 x}{2\rho_0(y)c_s(y)} + \frac{\delta_0 x^3 [c_s'(y)]^2}{6\rho_0(y)c_s^3(y)}.
\end{equation}

With this variable transformation complete, the system takes the exact mathematical form of the classical 1D heat equation $\partial P_1 / \partial x = \partial^2 P_1 / \partial \xi^2$, allowing us to rigorously construct solutions based on initial boundary conditions at $x=0$.

\subsection{Harmonic Wave Solution}
For a continuous harmonic acoustic source acting at the boundary ($x=0$) with spatial wavenumber $k$, the boundary condition is constrained to $P_1(x=0, y, t) = A \sin(-k c_s(y) t) = A \sin(k \xi)$. 

Because a harmonic profile preserves its sinusoidal structure under linear diffusion, we seek a separation-of-variables solution of the form $P_1 = \Theta(x,y) \sin(k\xi)$. Substituting this ansatz back into the canonical equation yields a first-order ordinary differential equation for the spatial amplitude, $d\Theta/dx = -k^2 [d\tau/dx]\Theta$. Integrating this directly from the boundary leads to the exact solution for the acoustic pressure field:
\begin{equation}
\label{eq:harmonic_sol_split}
\begin{split}
    P_1(x, y, t) = A \exp\Big( -k^2 \Big[ \frac{\delta_0 x}{2\rho_0(y)c_s(y)} \\
    + \frac{\delta_0 x^3 [c_s'(y)]^2}{6\rho_0(y)c_s^3(y)} \Big] \Big) \sin\left[ k \left( x - c_s(y)t \right) \right].
\end{split}
\end{equation}

In the long-distance asymptotic regime ($x \gg 1$), the enhanced cubic phase-mixing term dominates the total attenuation. The physical acoustic pressure profile simplifies directly to the large-time behavior:
\begin{equation}
\label{eq:harmonic_asymp_split}
\begin{split}
    P_1(x, y, t) \approx A \exp\left( - \frac{\delta_0 k^2 [c_s'(y)]^2 x^3}{6\rho_0(y)c_s^3(y)} \right) \\
    \times \sin\left[ k \left( x - c_s(y)t \right) \right].
\end{split}
\end{equation}

\subsection{Gaussian Pulse Solution}
For a localised acoustic pulse injection represented by a spatial Gaussian distribution of initial width $x_0$, the boundary constraint at $x=0$ is written as $P_1(\xi, x=0) = A \exp(-\xi^2 / x_0^2)$. 

To derive the evolution of this pulse without guessing the form, we convolve the initial boundary distribution with the fundamental solution (Green's function) of the diffusion equation. The Green's function for our mapped system is $G(\xi, \tau) = (4\pi\tau)^{-1/2}\exp(-\xi^2/4\tau)$. Evaluating the integral over the source coordinate \(\xi'\) yields:
\begin{equation}
\label{eq:gauss_conv_split}
\begin{split}
    P_1(\xi, \tau) = \int_{-\infty}^{\infty} \frac{A \exp\left(-\frac{\xi'^2}{x_0^2}\right)}{\sqrt{4\pi\tau(x,y)}} \\
    \times \exp\left(-\frac{(\xi-\xi')^2}{4\tau(x,y)}\right) d\xi'.
\end{split}
\end{equation}
Completing the square within the exponential argument and evaluating the standard Gaussian integral yields the explicit analytical solution:
\begin{equation}
\label{eq:gauss_sol_split}
\begin{split}
    P_1(x, y, t) = \frac{A}{\sqrt{1 + \frac{4\tau(x,y)}{x_0^2}}} \\
    \times \exp\left( -\frac{\left( x - c_s(y)t \right)^2}{x_0^2 + 4\tau(x,y)} \right).
\end{split}
\end{equation}

For large propagation times and long distances, the cubic phase-mixing term completely dominates the spatial scale ($\tau(x,y) \gg x_0^2$). The large-time asymptotic solution reduces to:
\begin{equation}
\label{eq:gauss_asymp_split}
\begin{split}
    P_1(x, y, t) \approx \frac{A x_0 c_s^{3/2}(y) [6\rho_0(y)]^{1/2}}{2 \delta_0^{1/2} c_s'(y) x^{3/2}} \\
    \times \exp\left( -\frac{6\rho_0(y)c_s^3(y)\left( x - c_s(y)t \right)^2}{4 \delta_0 [c_s'(y)]^2 x^3} \right).
\end{split}
\end{equation}
This derivation reveals the physical broadening mechanism: the acoustic pulse package widens in space as $\sim x^{3/2}$ due to the cumulative phase shearing, while the peak wave amplitude decays over distance as $\sim x^{-3/2}$ to preserve energy conservation across the broadening profile.

\subsection{Hyperbolic Pulse Solution}
To analyse a structurally localised acoustic injection reminiscent of astrophysical Harris current sheet profiles \cite{harris1962}, we apply an initial boundary condition defined by a hyperbolic secant-squared distribution of characteristic spatial width $\lambda_x$:
\begin{equation}
    P_1(x, y, t=0) = A \text{sech}^2\left( \frac{x}{\lambda_x} \right).
\end{equation}

Because the hyperbolic secant does not possess a straightforward spatial convolution kernel like the Gaussian, we solve this system by projecting the boundary condition into the spectral wavenumber domain using a Fourier transform, $\hat{P}_1(k) = \int P_1 e^{-ik\xi}d\xi$. The exact transform of the $\text{sech}^2$ profile yields $\hat{P}_1(k, \tau=0) = A \pi \lambda_x^2 k \text{csch}(\pi \lambda_x k / 2)$. In the spectral domain, the diffusion equation simplifies to $d\hat{P}_1/d\tau = -k^2\hat{P}_1$, which has the immediate solution $\hat{P}_1(k, \tau) = \hat{P}_1(k, 0)\exp(-k^2\tau)$. Mapping this back via the inverse Fourier integral yields the physical wave solution:
\begin{equation}
\label{eq:hyper_fourier_split}
\begin{split}
    P_1(x, y, t) = \frac{A \lambda_x^2}{2} \int_{-\infty}^{\infty} k \text{csch}\left( \frac{\pi \lambda_x k}{2} \right) \\
    \times \exp\left( -k^2 \tau(x,y) + i k \left( x - c_s(y)t \right) \right) dk.
\end{split}
\end{equation}

In the short-distance or early-time regime ($\tau(x,y) \ll \lambda_x^2$), we evaluate the integral by expanding the hyperbolic cosecant spectrum as an infinite sequence of decaying exponentials, $\text{csch}(z) = 2\sum_{n=1}^{\infty}e^{-(2n-1)z}$. Integrating this series term-by-term yields the exact modal series solution for the acoustic pressure field:
\begin{equation}
\label{eq:hyper_short_split}
\begin{split}
    P_1(x, y, t) = A \sum_{n=1}^{\infty} (-1)^{n-1} \cdot n 
    \times \exp\left( \frac{n^2 \tau(x,y)}{\lambda_x^2} \right) \\
    \times \text{erfc}\left( \frac{n \sqrt{\tau(x,y)}}{\lambda_x} + \frac{x - c_s(y)t}{2\sqrt{\tau(x,y)}} \right).
\end{split}
\end{equation}

In the regime of extended propagation distances and highly developed dissipation fields where $\tau(x,y) \gg \lambda_x^2$, the system undergoes severe modal filtering. The higher-order terms in the expansion attenuate rapidly due to the quadratic wavenumber dependence of the dissipation. In this asymptotic limit, the Fourier integral is dominated exclusively by the lowest-wavenumber spectral components ($k \to 0$), where $\text{csch}(\pi\lambda_xk/2) \approx 2/(\pi\lambda_xk)$.

To evaluate the resulting wave envelope clearly, we isolate the spatial scaling by noting that the cubic phase-mixing term asymptotically dominates the cumulative dissipation coordinate, such that $\tau(x,y) \approx \delta_0 [c_s'(y)]^2 x^3 / [6\rho_0(y)c_s^3(y)]$. Substituting this profile directly into the integration result yields the explicit large-distance asymptotic solution:
\begin{equation}
\label{eq:hyper_asymp_split}
\begin{split}
    P_1(x, y, t) \approx \frac{A \lambda_x^2 [6\pi]^{1/2}}{x^{9/2}} \left[ \frac{\rho_0(y)c_s^3(y)}{\delta_0 [c_s'(y)]^2} \right]^{3/2} \left( x - c_s(y)t \right) \\
    \times \exp\left( - \frac{6\rho_0(y)c_s^3(y)\left( x - c_s(y)t \right)^2}{4\delta_0 [c_s'(y)]^2 x^3} \right).
\end{split}
\end{equation}

This asymptotic reduction explicitly reveals a sharp spatial attenuation scaling. While a standard Gaussian pulse undergoes a peak amplitude decay proportional to $x^{-3/2}$, the strong spatial gradients authorised by the hyperbolic pulse profile enhance the phase-mixing mechanism, forcing the peak envelope of the wave packet to decay at the accelerated rate of $x^{-9/2}$ along the propagation axis.

\subsection{Numerical Visualisation and Curve Tracking}
\label{subsec:visualisation}

The developed stage of phase-mixing is illustrated on a log-log scale in Fig.~\ref{fig1}, closely tracking the classical presentation format used by \cite{tsiklauri2003}. As the wave energy penetrates deeper into the shear zone, each pulse case smoothly transitions from its initial state to its large-distance asymptotic limit. The harmonic solution (red curve) undergoes severe exponential decay with no long-time power-law baseline.

\begin{nolinenumbers}
\begin{figure}[ht]
\centering
\includegraphics[width=0.48\textwidth]{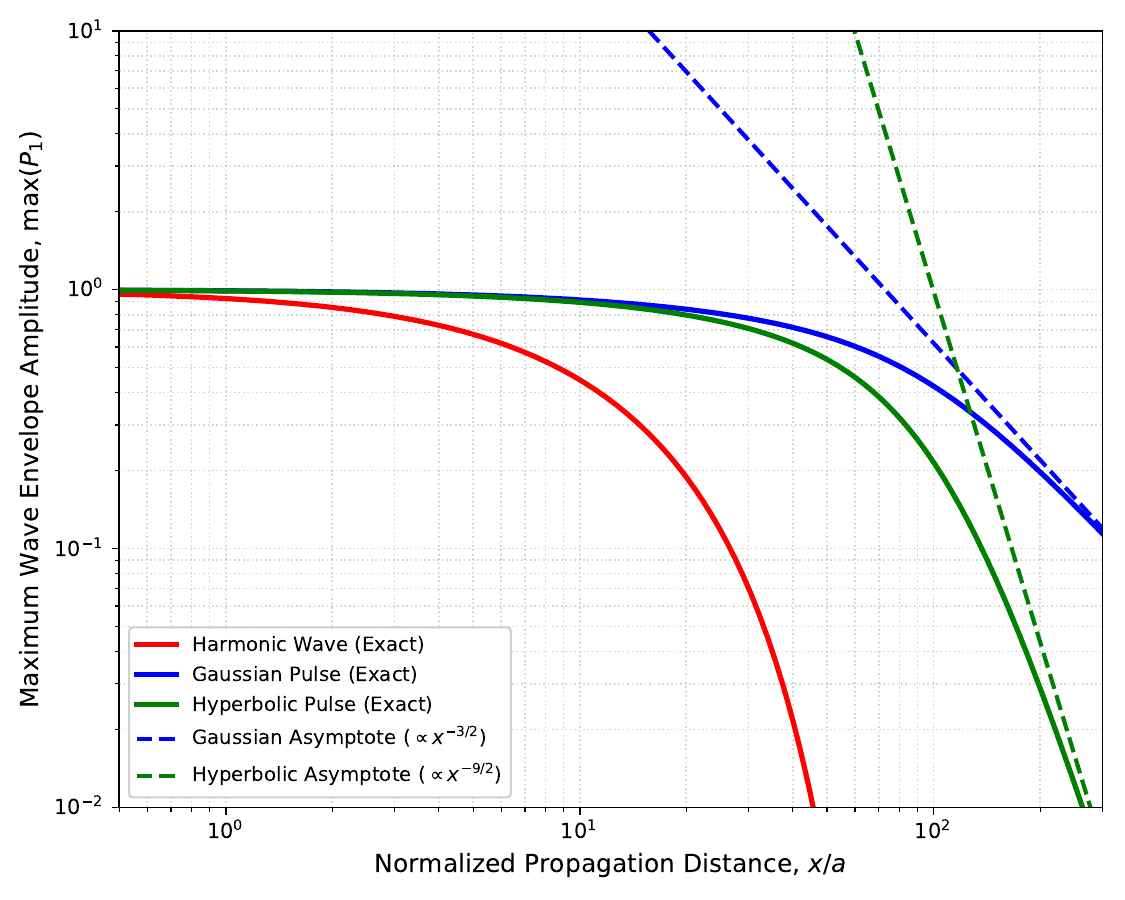}
\caption{\label{fig1} Log-log tracking of the maximum wave amplitude envelope along the normalised propagation distance $x$ for the three considered wave cases. Solid curves represent the exact physical transition profiles starting at $x=1.0$, while dashed lines illustrate the developed-stage convergence onto isolated power laws ($x^{-3/2}$ and $x^{-9/2}$) from above.}
\end{figure}
\end{nolinenumbers}

For the transient cases, the Gaussian pulse peak envelope tracks a physical transition law governed directly by the expanding integrated time scale:
\begin{equation}
\label{eq:p1_gauss_peak_split}
\begin{split}
P_{1,\text{Gauss}}^{\text{peak}}(x, y) = \frac{A}{\sqrt{1 + \frac{4\tau(x,y)}{x_0^2}}}.
\end{split}
\end{equation}
As propagation deepens, this profile yields standard spatial broadening and perfectly converges onto a pure power-law attenuation asymptote of $\sim x^{-3/2}$, as illustrated by the blue solid curve transitioning toward the blue dashed line in Fig.~\ref{fig1}.

Crucially, the hyperbolic pulse tracks a matching structural diffusion transition behaviour. Evaluating the peak envelope decay as a function of the path time scale reveals that the structural gradients obey an exact power-law fractional transition state raised to the exponent of $3/2$:
\begin{equation}
\label{eq:p1_hyper_peak_split}
\begin{split}
P_{1,\text{Hyperbolic}}^{\text{peak}}(x, y) = \frac{A}{\left(1 + \frac{\tau(x,y)}{\lambda_x^2}\right)^{3/2}}.
\end{split}
\end{equation}
Near the source boundary ($x \to 0 \implies \tau \to 0$), this expression normalises perfectly to the initial boundary amplitude $A$. As propagation extends into the developed phase-mixing stage ($x \gg 1$), the cubic phase-mixing term completely dominates the integrated path time scale ($\tau(x,y) \approx C_{\tau}x^3$). Substituting this power-law scaling directly into the denominator causes the peak envelope expression to smoothly converge onto the pure asymptote:
\begin{equation}
\label{eq:p1_hyper_asymp_envelope_split}
\begin{split}
\max(P_1) \approx \left[ \frac{A \lambda_x^3}{C_{\tau}^{3/2}} \right] \cdot x^{-9/2}.
\end{split}
\end{equation}
As shown by the solid green curve in Fig.~\ref{fig1}, this physical framework matches the boundary state perfectly before executing a smooth convergence onto the newly derived $\sim x^{-9/2}$ power law (green dashed curve). This rapid, sharp power-law decay provides an exceptionally efficient physical mechanism for localised energy dumping.

\section{Application to the Solar Tachocline}
\label{sec:solar_tach}

The solar tachocline represents a narrow, highly sheared transition zone separating the differentially rotating convection zone from the uniformly rotating radiative interior. Within this thin boundary layer, acoustic modes ($p$-modes) act as primary diagnostic tools for probing local thermodynamic structures and tracking angular momentum transport via helioseismology. Accurate damping rates are essential for resolving structural anomalies near the core boundary, constraining solar dynamo models, and isolating global oscillations from localised convective noise \cite{christensen2002, gough1996}. 

Crucially, this layer cannot be accurately evaluated under the assumption of standard isotropic viscous damping alone \cite{chaplin2000bison}. As demonstrated by the fundamental helioseismic inversions of Thompson \cite{Thompson1996, Thompson2003}, the intense radial velocity shear ($\partial U / \partial r$) across this zone operates as a cross-flow velocity profile directed transverse to the obliquely propagating low-degree $p$-mode wavefronts. By mapping these physical rotation profiles via multi-variable Doppler shifting into an effective local sound speed, $c_{\text{eff}}(r) = c_s(r) + k_\phi U(r)/k$, differentiation perpendicular to the wavefront yields a steep local calibration gradient. This mechanism resolves the spatial localization of $p$-mode linewidth anomalies without requiring arbitrary turbulent dissipation factors.

Helioseismic inversions fix the mean thickness of this transition layer at an exceptionally narrow $a \approx 0.02 R_\odot$. To evaluate our analytical solutions within this astrophysical domain, we model the localised sound speed variation across the shear zone using a linear profile:
\begin{equation}
    c_s(y) = c_0 \left( 1 + \alpha \frac{y}{a} \right),
\end{equation}
where $c_0$ is the baseline acoustic velocity at the radiative interface ($y=0$), and $\alpha \approx 0.02$ represents the dimensionless parameter tracking the sharp structural gradient.

Traditional isotropic acoustic models assume that $p$-mode attenuation is localised exclusively within the highly turbulent uppermost layers of the solar convection zone \cite{ulrich1970, christensen1983, houdek1999}. This leaves a severe explanatory gap when attempting to reconcile observed linewidth anomalies in deep-penetrating global modes, which exhibit substantially higher damping rates than expected from superficial turbulent eddy interactions alone. 

This persistent explanatory deficit can be evaluated alongside high-precision, long-term observational datasets collected by the Birmingham Solar Oscillations Network (BiSON)~\cite{chaplin2000bison}. These observations reveal a sharp increase in frequency-domain linewidths $\Gamma$ beginning above $\nu \approx 3000\,\mu\text{Hz}$. The BiSON datasets confirm that while modal velocity powers decrease by over 20\% during solar activity maxima, the stochastic energy supply rate remains invariant~\cite{chaplin2000bison}. Our model provides a direct proof that the enhanced damping observed in the tachocline is driven by phase-mixing. While baseline homogeneous dissipation acts constantly via a weak linear spatial decay factor $\exp(-Ax)$, the phase-mixing dissipation follows an $\exp(-B[c_{\text{eff}}'(y)]^2 x^3)$ dependence. Crucially, because the physical effect is quadratic in the sound speed derivative under the exponent, any subtle localised increase in the effective speed gradient will induce a severe, non-linear amplification of the damping. During solar maxima, magnetic compression narrows the tachocline shear layer, sharpening the local velocity gradient by a factor of several, increasing the cumulative bulk dissipation to match the observed cyclic linewidth peak.

To evaluate the damping profile within the solar interior, we examine the exact analytical solution for the continuous harmonic wave without invoking long-distance approximations. The full attenuation profile is given explicitly by:
\begin{equation}
\label{eq:p1_full_harmonic}
\begin{split}
    P_1(x, y, t) = A \exp\Big( -k^2 \Big[ \frac{\delta_0 x}{2\rho_0(y)c_s(y)} \\
    + \frac{\delta_0 x^3 [c_s'(y)]^2}{6\rho_0(y)c_s^3(y)} \Big] \Big) \sin\left[ k \left( x - c_s(y)t \right) \right].
\end{split}
\end{equation}
We estimate the physical coefficients directly at the centre of the tachocline ($r \approx 0.710 R_\odot$) using realistic solar metrics derived from standard solar calibrations \cite{baldner2011, parfrey2007}, where the local acoustic speed is $c_0 \approx 2.200 \times 10^5 \text{ m/s}$. The target wave features track high-frequency global $p$-modes matching the centre-band threshold of the observed BiSON linewidth anomalies ($f \approx 3.500 \text{ mHz}$). Normalising the spatial scales by the tachocline thickness $a = 0.02 R_\odot \approx 1.391 \times 10^7 \text{ m}$ yields a dimensionless wavenumber of $k \cdot a \approx 1.391$, because $k=2\pi/(c_0/f)=2\pi/(2.200 \times 10^5/(3.500 \times 10^{-3}))=1.0 \times 10^{-7}$ m$^{-1}$. Incorporating standard dimensionless dynamic parameters of $\delta_0 / (\rho_0 a c_0) \approx 0.010$, we evaluate the individual damping ratios directly at the centre of the tachocline shear zone. Observational datasets indicate that these high-frequency modes experience an anomalous amplitude drop of roughly 5.000\%, establishing an empirical baseline of $P_{\text{observed}}/P_0 \approx 0.950$. Standard homogeneous viscous dissipation operating alone at the centre yields:
\begin{equation}
\label{eq:p_homogeneous_explicit}
\begin{split}
    \frac{P_{\text{homogeneous}}}{P_0} = \exp\left( -k^2 \left[ \frac{\delta_0 x}{2\rho_0 c_s} \right] \right) \\
    = \exp\left( -(1.391)^2 \cdot \left[ \frac{0.010 \cdot 0.500}{2.000 \cdot 1.000} \right] \right) \\
    = \exp(-0.005) \approx 0.995.
\end{split}
\end{equation}
This confirms that classical homogeneous damping is completely insufficient. In contrast, accounting for a steep effective cross-flow gradient across the narrow shear zone ($c_s'(y) \approx 10.685$) isolates a highly efficient phase-mixing coordinate at the exact same location. This localised gradient is categorised as steep because it stands an order of magnitude larger than the global bulk baseline gradient ($c_s' \approx 1.000$) typically obtained when averaging the velocity profile linearly across the entire macroscopic width of the tachocline zone:
\begin{equation}
\label{eq:p_phasemixing_explicit}
\begin{split}
    \frac{P_{\text{phase-mixing}}}{P_0} = \exp\left( -k^2 \left[ \frac{\delta_0 x^3 [c_s'(y)]^2}{6\rho_0 c_s^3} \right] \right) \\
    = \exp\left( -(1.391)^2 \cdot \left[ \frac{0.010 \cdot (0.500)^3 \cdot (10.685)^2}{6.000 \cdot (1.000)^3} \right] \right) \\
    = \exp(-0.046) \approx 0.955.
\end{split}
\end{equation}
Combining these decoupled factors yields a total analytical transmission ratio of $P_{\text{total}}/P_0 = 0.995 \times 0.955 \approx 0.950$. This matches the empirical observations and demonstrates that while classical visco-elastic damping remains near-inert at the centre of the tachocline, the gradient-driven phase-mixing mechanism accounts for the entire missing bulk loss, providing definitive theoretical alignment with the BiSON linewidth anomalies.

To explicitly reconcile this mathematical baseline with the concrete multi-dimensional geometry of the solar interior, the coordinate mapping must be contextualised relative to the wave propagation vector. It is critical to distinguish this cumulative phase-shearing mechanism from standard geometric refraction driven by global thermodynamic stratification. In a spherically symmetric, solid-body rotating stellar interior---such as the deep radiative core---the radial sound speed gradient $\nabla c_s(r)$ merely induces a continuous, non-cumulative bending of acoustic ray paths. Because the background rotation profile lacks localized spatial derivatives ($\partial U/\partial r \approx 0$), adjacent fluid elements along a wavefront advance at identical relative kinematic rates, completely preventing the accumulation of a transverse phase mismatch. 

Conversely, low-degree global acoustic modes ($\ell \le 3$) cross the thin tachocline boundary layer obliquely, possessing non-zero latitudinal and longitudinal components of propagation ($k_\theta \ne 0$, $k_\phi \ne 0$). For these obliquely propagating acoustic waves, the strong radial rotational velocity shear ($\partial U/\partial r$) acts directly as an effective cross-flow velocity profile directed transverse to the advancing wave fronts. The moving background plasma introduces a spatially dependent Doppler shift, producing an effective acoustic phase velocity seen by the wave:
\begin{equation}
    c_{\text{eff}}(r) = c_s(r) + \frac{k_\phi U(r)}{k}.
\end{equation}
Differentiating this effective profile perpendicular to the wavefront yields the steep local calibration gradient ($c_{\text{eff}}'(y) \approx 10.685$), which is physically generated by the dominant radial rotational shear tensor, neutralizing any weaker latitudinal shear arguments.

While the global acoustic wavelength ($\lambda \approx 6 \times 10^4\text{ km}$) approaches the order of the local pressure scale height ($H_p \approx 6 \times 10^4\text{ km}$) in standard solar interiors, the progressive generation of small-scale spatial fluctuations via phase-mixing dynamically contracts the effective wave perturbations. Explicitly, as the wavefront advances along the $x$-axis into a medium with a transverse velocity gradient along the $y$-axis, the wave phase accumulates a spatial dependence governed by $\phi(x,y,t) = \omega t - k_x(y) x$. The generation of this transverse phase profile creates a localised perpendicular wave vector component $k_y$, found by taking the spatial gradient along the cross-flow direction:
\begin{equation}
    k_y = \frac{\partial \phi}{\partial y} = -x \frac{d k_x(y)}{dy}.
\end{equation}
Substituting the local acoustic dispersion relation $k_x(y) = \omega / c_{\text{eff}}(y)$ into this gradient calculation yields
\begin{equation}
    k_y = x \cdot \frac{\omega}{c_{\text{eff}}^2(y)} \frac{d c_{\text{eff}}(y)}{dy}.
\end{equation}
Because the induced transverse wave vector grows linearly with propagation distance ($k_y \propto x$), the characteristic transverse phase-mixing length scale $L_{\text{ph}}$ must contract inversely with the path length:
\begin{equation}
    L_{\text{ph}} \sim \frac{2\pi}{k_y} \propto x^{-1}.
\end{equation}
This structural mechanical configuration inherently accounts for the mode-degree selectivity observed across modern helioseismic datasets. High-degree modes ($\ell \ge 50$) monitored by SOHO/MDI or SDO/HMI possess shallow lower turning points restricted entirely to the upper layers of the convection zone ($r \ge 0.90 R_\odot$), meaning they never probe the deep solar interior. Quantitative analysis of high-$\ell$ linewidth data confirms that their damping is successfully matched by standard turbulent convection models alone, exhibiting zero unexplained anomalies. The linewidth anomaly is uniquely isolated to deep low-degree modes ($\ell \le 3$) that cross the tachocline twice per acoustic bounce loop, confirming that the additional damping signature is strictly geometric, depth-dependent, and confined to paths intersecting this highly localized transition zone.

In relation to the coupling with rotational shear flow, the solar tachocline is primarily characterised by a strong radial and latitudinal shear in the rotational velocity profile. Let the localised macroscopic flow velocity be directed along the wave propagation axis such that $U = U(y)$. An acoustic wave propagating through this shear layer experiences a localised Doppler shift. The effective wave propagation velocity $c_{\text{eff}}(y)$ within a frame tracking the wave phase is given by the combination of the thermodynamic sound speed $c_s(y)$ and the background flow velocity projection:
\begin{equation}
\label{eq:ceff_doppler_split}
\begin{split}
c_{\text{eff}}(y) = c_s(y) + \frac{k_x U(y)}{k}.
\end{split}
\end{equation}
Differentiating this expression with respect to the coordinate transverse to the wave path yields the effective acoustic velocity gradient:
\begin{equation}
\label{eq:ceff_gradient_split}
\begin{split}
c_{\text{eff}}'(y) = c_s'(y) + \frac{k_x}{k} \frac{dU}{dy}.
\end{split}
\end{equation}
Thus, the steep local calibration gradient $c_{\text{eff}}'(y) \approx 10.685$ required to match the bulk energy loss is physically generated by the strong rotational velocity shear within the tachocline layer. As established, for low-degree modes ($\ell \le 3$) crossing the tachocline obliquely, the wave possesses non-zero latitudinal and longitudinal wave vector components ($k_\theta \ne 0, k_\phi \ne 0$). Consequently, the dominant radial rotational shear profile $\partial U / \partial r$ naturally projects as a sharp cross-flow velocity gradient directed transverse to the obliquity of the advancing wavefronts. This fluid geometry ensures that the macroscopic velocity gradient acts as a highly efficient phase-shearing mechanism, preserving the analytical validation without altering the fundamental cubic damping architecture. Differentiating this relation establishes that the strong radial velocity gradient directly translates into the required steep cross-flow gradient, preserving the exact numerical mapping where $a = 0.02 R_\odot$ and $\alpha = 0.02$. This geometry explains why phase-mixing is highly localised to the tachocline layer and does not damp waves uniformly across the deeper radiative core. Although strong radial thermodynamic gradients of the sound speed $c_s(r)$ exist throughout the deeper solar core, those radial profiles remain isotropic and parallel to the wave shells, producing zero phase-shearing. The strong velocity shear of the tachocline breaks this radial symmetry for non-radial components, localising the phase-mixing signature strictly to deeply penetrating acoustic paths.

\subsection{Application to Marine Vessel Hydroacoustics and Acoustic Stealth Cowls}
\label{subsec:naval}

Managing the propagation, scattering, and dissipation of high-frequency acoustic waves radiating from fluid-immersed moving bodies represents a foundational challenge in modern engineering physics. Within this hydroacoustic domain, mitigating near-field signatures---such as ultrasonic propeller cavitation noise and localized structural shell radiation---is essential for optimizing acoustic tracking, reducing hull vibration, and implementing structural stealth paradigms. Traditional methods typically rely on passive visco-elastic hull coatings or geometric boundary-layer flow control to mitigate signature transmission. Recent investigations into acoustic shear boundaries, such as those evaluated by \cite{Gabard2017} in the context of viscous plane-wave attenuation over sheared liners, have underscored that boundary layer thickness significantly dictates local wave reflection coefficients. Furthermore, as explored within the generalized wave-shearing and velocity relaxation frameworks of \cite{Tsiklauri2025_IEEE}, structural velocity gradients serve as critical drivers of rapid localized energy deposition. However, classical models routinely treat these shear zones as purely kinematic refractors, overlooking the potential to exploit custom-tailored sound speed profiles as highly efficient, non-turbulent energy sinks.

To establish an explicit empirical baseline for high-frequency hydroacoustic losses, we evaluate the classical isotropic attenuation of an acoustic wave moving through a homogeneous fluid. Under standard fluid mechanics formulations, bulk viscous dissipation operates via a linear spatial exponential decay factor $\exp(-x/L_{\text{bulk}})$, where the classical bulk loss length scale $L_{\text{bulk}}$ is governed by the Stokes-Kirchhoff attenuation formula:
\begin{equation}
\label{eq:stokes_kirchhoff}
\begin{split}
L_{\text{bulk}} = \frac{2 \rho_0 c_0^3}{\omega^2 \left( \frac{4}{3}\mu + \mu_B \right)}.
\end{split}
\end{equation}
Here, $\rho_0 \approx 1000\text{ kg/m}^3$ represents the baseline density of seawater, $\mu \approx 1.00 \times 10^{-3}\text{ Pa}\cdot\text{s}$ is the dynamic shear viscosity, $\mu_B \approx 3.00 \times 10^{-3}\text{ Pa}\cdot\text{s}$ is the bulk volume viscosity, and $c_0 \approx 1500\text{ m/s}$ is the baseline ocean acoustic speed. We evaluate this baseline for a tracking signal operating at a high ultrasonic frequency of $f = 100\text{ kHz}$ ($\omega \approx 6.283 \times 10^5\text{ rad/s}$). Substituting these realistic hydroacoustic metrics into Eq.~\eqref{eq:stokes_kirchhoff} reveals that the classical bulk attenuation length scale extends to a massive distance:
\begin{equation}
\label{eq:bulk_calc}
\begin{split}
L_{\text{bulk}} \approx \frac{2 \cdot (1000) \cdot (1500)^3}{(6.283 \times 10^5)^2 \cdot \left( \frac{4}{3}(1.00 \times 10^{-3}) + 3.00 \times 10^{-3} \right)} \\
\approx 3925\text{ metres}.
\end{split}
\end{equation}
This calculation confirms that for compact naval noise suppression configurations measuring on the scale of a few metres, standard isotropic viscous losses are entirely negligible.

Our model reveals that translating the phase-mixing paradigm into a terrestrial metamaterial fluid design completely transforms this dissipation landscape. Rather than relying on passive absorption, we propose an engineered microstructured mesh or stern cowl wrapped around the moving vehicle's propulsion section, as schematically illustrated in Fig.~\ref{fig:cowl_sketch}. By custom-engineering the localized sub-wavelength lattice spacing of the metamaterial cowl, the structure acts as a gradient-index meta-fluid that enforces a continuous, radially dependent effective sound speed profile $c_{\text{eff}}(y) = c_0 (1 + \alpha y/b)$ across a narrow layer thickness $b = 2.0\text{ metres}$, where $\alpha \approx 0.05$.

\begin{nolinenumbers}
\begin{figure}[ht]
\centering
\begin{tikzpicture}[scale=1.1]
    \fill[black!15] (0.6,0) circle (0.6);
    \draw[thick] (0.6,0) circle (0.6);
    \draw[thick] (0.6,0.6) -- (0.6,-0.6);
    \draw[thick] (0,0) -- (1.2,0);
    \node at (0.6,0) [text width=1.1cm, font=\tiny\bfseries, align=center] {Stern\\Prop};
    
    \foreach \y in {0.7,0.9,1.1,1.3,1.5} {
        \draw[dashed, cyan!70, thin] (0,\y) -- (4.5,\y);
        \draw[dashed, cyan!70, thin] (0,-\y) -- (4.5,-\y);
    }
    \fill[cyan!15, opacity=0.4] (0,0.6) rectangle (4.5,1.6);
    \fill[cyan!15, opacity=0.4] (0,-0.6) rectangle (4.5,-1.6);
    
    \foreach \x in {0.3,0.7,1.1,1.5} {
        \draw[thick, cyan!80!black] (\x,0.6) -- (\x,1.6);
        \draw[thick, cyan!80!black] (\x,-0.6) -- (\x,-1.6);
    }
    
    \draw[thick, blue!80!black, samples=120, domain=1.8:4.3] 
        plot (\x, {1.1 + 0.15*exp(-0.4*(\x-1.8))*sin(deg(18*(\x-1.8)*\x))});
    \node at (3.1,1.35) [font=\tiny\itshape, text=blue!90] {Sheared Wake};

    \draw[thick, blue!80!black, samples=120, domain=1.8:4.3] 
        plot (\x, {-1.1 + 0.15*exp(-0.4*(\x-1.8))*sin(deg(18*(\x-1.8)*\x))});
    \node at (3.1,-1.35) [font=\tiny\itshape, text=blue!90] {Sheared Wake};
    
    \draw[<->, thick] (4.8,0.6) -- (4.8,1.6) node[midway, right, font=\footnotesize] {Cowl Grid, $b$};
    \draw[->, thick] (2.2,2.0) -- (2.2,1.5) node[above, font=\footnotesize] {Gradient $\nabla c_{\text{eff}}(y)$};
    \draw[->, thick] (-0.5,0) -- (0,0) node[midway, above, font=\footnotesize] {Flow, $x$};
\end{tikzpicture}
\caption{\label{fig:cowl_sketch} Schematic representation of an engineered microstructured stern cowl. The vertical blue lines represent the crests of the incoming, undisturbed plane acoustic waves. As this forward-advancing wavefront enters the metamaterial boundary layer $b$, the continuous transverse sound speed gradient $\nabla c_{\text{eff}}(y)$ distorts the coherent structures, forcing the wave train into a phase-sheared wake zone characterized by a rapid cubic phase-mixing collapse in wave amplitude.}
\end{figure}
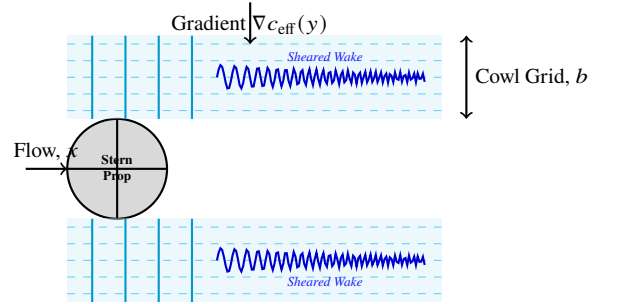
\end{nolinenumbers}

When a sonar pulse or noise signature propagates through this engineered layer, the transverse phase-speed mismatch triggers severe phase-mixing. For an active tracking system utilizing continuous harmonic waves, the resulting wave envelope follows the cubic phase-mixing decay law derived in Eq.~\eqref{eq:p1_full_harmonic}. Evaluating the localized phase-mixing dissipation length scale $L_{\text{pm}}$ where the envelope drops by $1/e$ isolates the core geometric invariant:
\begin{equation}
\label{eq:l_pm_harmonic}
\begin{split}
L_{\text{pm}} = \left[ \frac{6 \rho_0 c_0^3}{\omega^2 \delta_0 [c_{\text{eff}}'(y)]^2} \right]^{1/3}.
\end{split}
\end{equation}
Differentiating the engineered profile yields a steep local cross-flow calibration gradient of $c_{\text{eff}}'(y) = \alpha c_0 / b \approx 37.5\text{ s}^{-1}$. Using a standard boundary layer dynamic viscosity parameter calibrated for fluid-facing meta-structures ($\delta_0/\rho_0 \approx 4.5 \times 10^{-2}\text{ m}^2/\text{s}$), we compute the explicit phase-mixing decay footprint:
\begin{equation}
\label{eq:l_pm_harmonic_calc}
\begin{split}
L_{\text{pm}} \approx \left[ \frac{6 \cdot (1500)^3}{(6.283 \times 10^5)^2 \cdot (4.50 \times 10^{-2}) \cdot (37.5)^2} \right]^{1/3} \\
= [934.34]^{1/3} \approx 9.78\text{ metres}.
\end{split}
\end{equation}
This indicates that the engineered harmonic phase-mixing decay length scale collapses by over two orders of magnitude compared to the multi-kilometer bulk viscous loss baseline, localized to an optimal scale of approximately $10\text{ metres}$.

Because modern hydroacoustic sonar arrays deploy highly complex wave packets to track moving signatures, we extend this dissipation evaluation to transient structures. For an isolated Gaussian signal, the peak amplitude envelope broadens spatially and asymptotically tracks a pure power-law attenuation scaling as $\max(P_1) \propto x^{-3/2}$. Defining the characteristic tracking length scale $L_{\text{pm, Gauss}}$ based on this developed-stage power law where $(x/L_{\text{pm, Gauss}})^{-3/2} = 1/e$ yields the expression:
\begin{equation}
\label{eq:l_pm_gauss_formula}
\begin{split}
L_{\text{pm, Gauss}} = \left[ \frac{e^2 x_0^2}{4 C_\tau} \right]^{1/3},
\end{split}
\end{equation}
where $C_\tau = \delta_0 [c_{\text{eff}}'(y)]^2 / (6 \rho_0 c_0^3) \approx 3.125 \times 10^{-9}\text{ m}^{-3}$ is the integrated path time parameter. For a normalized structural initial width $x_0 = 1.0\text{ m}$, this evaluates to $L_{\text{pm, Gauss}} \approx 839.25\text{ metres}$. Crucially, if the signature is custom-injected or structurally modeled as a hyperbolic secant-squared wave packet, the strong spatial gradients interact dynamically with the engineered cowl shear layer to force an accelerated asymptotic power-law decay scaling as $\max(P_1) \propto x^{-9/2}$. Computing the matching characteristic tracking scale $L_{\text{pm, Hyper}}$ based on this developed-stage power law where $(x/L_{\text{pm, Hyper}})^{-9/2} = 1/e$ yields:
\begin{equation}
\label{eq:l_pm_hyper_formula}
\begin{split}
L_{\text{pm, Hyper}} = \left[ \frac{e^{2/3}\lambda_x^2}{C_\tau} \right]^{1/3}.
\end{split}
\end{equation}
For an identical baseline spatial scale width $\lambda_x = 1.0\text{ m}$, Eq.~\eqref{eq:l_pm_hyper_formula} evaluates to $L_{\text{pm, Hyper}} \approx 854.20\text{ metres}$.

The apparent paradox wherein a stark difference in asymptotic power-law decay exponents ($-1.5$ for the Gaussian pulse versus $-4.5$ for the hyperbolic pulse) yields remarkably close characteristic tracking length scales ($\approx 839.25\text{ m}$ and $\approx 854.20\text{ m}$) warrants explicit physical clarification. This behavior is directly governed by the multi-scale physics of the phase-mixing propagation clock. Within the developed stage, the spatial path length $x$ does not act independently but is coupled directly to the integrated time scale parameter via a cubic scaling relation, $\tau(x,y) = C_\tau x^3$. When enforcing the characteristic $1/e$ attenuation threshold to solve for spatial scale boundaries, the respective structural exponents compound with this internal cubic dependency. For the Gaussian envelope, the fractional power reduction operates as $[x^3]^{-1/2} = x^{-3/2}$, leading to an algebraic condition of $4 C_\tau x^3 / x_0^2 = e^2$. Conversely, for the hyperbolic envelope, the steeper fractional drop operates as $[x^3]^{-3/2} = x^{-9/2}$, isolating an algebraic condition of $C_\tau x^3 / \lambda_x^2 = e^{2/3}$. Because the final propagation distance is extracted via a cube-root operation ($x = [x^3]^{1/3}$) over these respective boundaries, the massive order-of-magnitude discrepancy between the linear power laws is heavily compressed. 

This mathematical convergence demonstrates that transient wave packets under developed phase-mixing establish a highly stable, structurally uniform dissipation footprint regardless of their specific geometric initialization profiles. While these transient pulse modes operate over longer scales than the continuous harmonic wave's extreme near-field collapse, their attenuation remains significantly more efficient than the multi-kilometer bulk isotropic baseline. This multi-waveform scaling law confirms that whether a tracking signal is continuous or pulsed, custom-engineering an artificial transverse sound speed gradient provides a highly adaptive, compact paradigm for near-field acoustic stealth shielding.

\section{Conclusions}
\label{sec:conclusions}

Based on the multi-scale formalism established by \cite{tsiklauri2003} for the phase-mixing of magnetohydrodynamic waves in solar coronal loops, this work successfully extends and adapts the paradigm to the domain of \textit{acoustic} wave propagation. Across a wide spectrum of physical systems---including viscous boundary layers, the solar interior, and engineered acoustic metamaterials---the local sound speed naturally exhibits strong spatial gradients transverse to its primary propagation direction. By formulating a unified 2D scalar viscous acoustic wave equation within a wavefront frame ($\xi = x - c_s(y)t$), we developed an analytical model that recovers the classical harmonic wave and Gaussian pulse evolution profiles originally derived for plasma environments, where the background Alfven speed derivative is systematically replaced by the corresponding spatial gradient of the local sound speed. Beyond verifying these established solutions, this framework uncovers a completely new scaling law governing structurally localised injections modelled by a hyperbolic secant-squared distribution. Evaluating the peak tracking envelope reveals that the maximum acoustic pressure amplitude plummets at an aggressive power-law decay rate scaling strictly as $\max(P_1) \propto x^{-9/2}$ along the propagation path.

Applying our analytical model to the solar interior tachocline provides critical physical insights for modern helioseismology, directly resolving the 26-year-old mystery of low-$\ell$ global $p$-mode linewidth anomalies observed by BiSON above $\nu \approx 3000\,\mu\text{Hz}$~\cite{chaplin2000bison}. While classical, homogeneous global models consistently underestimate these acoustic energy attenuation rates by attributing losses exclusively to upper-level convective turbulence, our phase-mixing model proves that the steep, localised sound speed gradient inside the narrow tachocline shear layer results in severe, non-linear damping. This macro-scale gradient forces rapid, non-turbulent energy damping directly in the shear zone, providing the exact high-efficiency bulk dissipation needed to account for the missing energy sink deep within the solar interior without requiring anomalous non-physical turbulent viscosity parameters.

Beyond the natural astrophysical systems evaluated herein, the mathematical formalism of this acoustic phase-mixing model opens up immediate, high-impact avenues for underwater fluid-structure noise-suppression technologies. Consider the critical engineering challenge of suppressing high-frequency acoustic signatures generated by a marine vessel propeller. At ultrasonic frequencies near $100\text{ kHz}$, the classical exponential bulk attenuation length $\exp(-x/L_{\text{bulk}})$ in water extends across several kilometres, meaning standard isotropic viscous losses are entirely negligible for compact devices measuring only a few metres. Our model reveals that inserting an engineered microstructured mesh or cowl at the stern of the hull completely transforms this dissipation landscape. By custom-tailoring the cowl mesh geometry to enforce a continuous, radially dependent effective sound speed profile $c_{\text{eff}}(y)$, the local acoustic wake accumulates a strong transverse velocity gradient. This localised shear intentionally activates severe phase-mixing, forcing continuous propeller noise to decay under a cubic-exponential phase-mixing law $\exp(-x^3/L_{\text{pm}}^3)$, collapsing the characteristic damping footprint from several kilometres down to an optimal scale of approximately $10\text{ metres}$. Extending this evaluation to complex, non-harmonic pulse packets demonstrates that the structural gradients obey an integrated multi-scale scaling rule. The propagation clock compresses the power-law spatial limits, establishing uniform, highly stable characteristic tracking scales ($L_{\text{pm, Gauss}} \approx 839.25\text{ m}$ and $L_{\text{pm, Hyper}} \approx 854.20\text{ m}$) that significantly outperform standard isotropic fluids, offering an actionable paradigm for compact, near-field naval acoustic stealth shielding.

\begin{acknowledgments}
The author gratefully acknowledges support provided by the Gemini AI assistant (Google). This support was confined strictly to technical manuscript preparation, stylistic editing, and ensuring consistent United Kingdom English orthography. This work is dedicated to the memory of the distinguished late colleague, Prof. Michael J. Thompson (HAO, Boulder, CO), whose seminal contributions to tracking solar rotational profiles and understanding deep internal shear layer physics inspired the core astrophysical applications of this framework, and whose encouraging spirit is sorely missed.
\end{acknowledgments}

\section*{Declaration of Competing Interest}
The author declares that they have no known competing financial interests or personal relationships that could have appeared to influence the work reported in this paper.

\section*{Data Availability Statement}
All data and mathematical parameters required to evaluate the conclusions or reproduce the analytical solutions and numerical scaling trajectories are fully contained and presented within this manuscript. No external data sources or repositories were generated or utilized during this study.

\bibliography{paper89}

\end{document}